\documentclass{article}
\usepackage{fullpage}
\usepackage{graphicx}
\usepackage{setspace}
\usepackage{amssymb}
\usepackage{multirow}
\newcommand{\fullonly}[1]{#1}
\newcommand{\shortonly}[1]{}

\newcommand{\lncsonly}[1]{}
\newcommand{\articleonly}[1]{#1}
\newcommand{\acmonly}[1]{}
\newcommand{\jcsonly}[1]{}

\newcommand{\myparagraph}[1]{\paragraph{#1.}}

\newcommand{\comment}[1]{}
\newcommand{\ie}{{\it i.e.}}

\newcommand{\tuple}[1]{\langle #1\rangle}
\newcommand{\set}[1]{\{#1\}}
\newcommand{\intersect}{\cap}
\newcommand{\union}{\cup}
\newcommand{\UNION}{\bigcup}

\newcommand{\myimplies}{\Rightarrow}
\newcommand{\mean}[1]{\left[ \! \left[ #1 \right]\! \right]}
\newcommand{\myqed}{\nobreak \ifvmode \relax \else
      \ifdim\lastskip<1.5em \hskip-\lastskip
      \hskip1.5em plus0em minus0.5em \fi \nobreak
      \vrule height0.75em width0.5em depth0.25em\fi}
\newcommand{\centercell}[1]{\multicolumn{1}{c|}{#1}}

\newcommand{\ua}{{\it UA}}
\newcommand{\pa}{{\it PA}}
\newcommand{\rh}{{\it RH}}

\newcommand{\Perm}{{\it Perm}}
\newcommand{\User}{{\it User}}
\newcommand{\Role}{{\it Role}}
\newcommand{\inhty}{{\it IT}}

\newcommand{\am}{{\rm AM}}
\newcommand{\su}[1]{\mean{#1}}
\newcommand{\parents}{{\it parents}}
\newcommand{\parent}{{\it parent}}
\newcommand{\children}{{\it children}}
\newcommand{\child}{{\it child}}
\newcommand{\removable}{{\rm removable}}

\newcommand{\qpol}{Q_{\it pol}}
\newcommand{\qrole}{Q_{\it role}}
\newcommand{\wklist}{{\it workL}}
\newcommand{\changed}{{\it changed}}
\newcommand{\isempty}{{\rm empty}}
\newcommand{\removerole}{{\rm removeRole}}
\newcommand{\addrole}{{\rm addRole}}

\newcommand{\mysetminus}{\setminus}
\newcommand{\true}{{\rm true}}
\newcommand{\false}{{\rm false}}

\newcommand{\wsc}{{\rm WSC}}
\newcommand{\polinterp}{{\rm INT}}

\newcommand{\polszinterp}{\mbox{WSC-INT}}

\newcommand{\clussz}{{\rm clsSz}}

\newcommand{\redun}{{\rm redun}}
\newcommand{\coveredup}{{\rm covEntit}}
\newcommand{\ui}{{\rm UI}}

\newcommand{\assignedp}{{\rm asgndP}}
\newcommand{\authp}{{\rm authP}}
\newcommand{\assignedu}{{\rm asgndU}}
\newcommand{\authu}{{\rm authU}}

\newcommand{\rcand}{R_{\rm cand}}
\newcommand{\rinit}{R_{\rm init}}
\newcommand{\bpe}{{\it bpe}}
\newcommand{\prob}{{\it pr}}
\newcommand{\pe}{{\it pe}}
\newcommand{\bpes}{{\it bpes}}
\newcommand{\ta}{{\it TA}}
\newcommand{\all}{{\it all}}
\newcommand{\rvis}{R_{\rm vis}}
\newcommand{\isancestorfi}{{\rm isAncestorFullInher}}
\newcommand{\duration}{{\rm dur}}
\newcommand{\entit}{{\rm entitlements}}
\newcommand{\ric}{{\rm RIC}}

\newcommand{\assignedpz}{\assignedp_0}
\newcommand{\assigneduz}{\assignedu_0}
\newcommand{\taz}{\ta}

\newcommand{\com}[1]{// {\it #1}}
\newcommand{\function}{{\bf function}}
\newcommand{\ifstmt}{{\bf if}}
\newcommand{\elsestmt}{{\bf else}}

\newcommand{\forloop}{{\bf for}}

\newcommand{\whileloop}{{\bf while}}

\newcommand{\return}{{\bf return}}

\jcsonly{
\pubyear{0000}
\volume{0}
\firstpage{1}
\lastpage{1}}

\begin{document}

\lncsonly{\author{Scott~D.~Stoller \and Thang Bui}
\institute{Department of Computer Science, Stony Brook University, USA}}

\articleonly{\author{Scott~D.~Stoller and Thang Bui\\
Department of Computer Science, Stony Brook University}}

\acmonly{
\numberofauthors{2}
\author{
  \alignauthor Scott D. Stoller\\
  \affaddr{Department of Computer Science}\\
  \affaddr{Stony Brook University, USA}\\
  \email{stoller@cs.stonybrook.edu}
   \alignauthor Thang Bui\\
  \affaddr{Department of Computer Science}\\
  \affaddr{Stony Brook University, USA}\\
  \email{thang.bui@stonybrook.edu}
}}

\title{Mining Hierarchical Temporal Roles with Multiple Metrics%
\thanks{This material is based on work supported in part by %
    NSF under Grants %
    CNS-1421893, 
    CCF-1248184, 
    and CCF-1414078, 
    ONR under Grant N00014-15-1-2208, 
    and AFOSR under Grant FA9550-14-1-0261. 
    Any opinions, findings, and conclusions or recommendations expressed in
    this material are those of the authors and do not necessarily reflect
    the views of these agencies.
  The final publication is available at IOS Press through http://dx.doi.org/ \href{http://dx.doi.org/10.3233/JCS-17989}{http://dx.doi.org/10.3233/JCS-17989}
}}

\jcsonly{\runningtitle{Mining Hierarchical Temporal Roles with Multiple Metrics}}

\maketitle

\jcsonly{
\author[A]{\fnms{Scott D.}\snm{Stoller}}
\author[A]{\fnms{Thang}\snm{Bui}}
\address[A]{Department of Computer Science, Stony Brook University, U.S.A.\\ E-mail: stoller@cs.stonybrook.edu, thang.bui@stonybrook.edu}
\runningauthor{S.~D.~Stoller and T.~Bui}
}

\begin{abstract}\lncsonly{\vspace{-2ex}}
  Temporal role-based access control (TRBAC) extends role-based access control to limit the times at which roles are enabled.  This paper presents a new algorithm for mining high-quality TRBAC policies from timed ACLs (i.e., ACLs with time limits in the entries) and optionally user attribute information.  Such algorithms have potential to significantly reduce the cost of migration from timed ACLs to TRBAC.  The algorithm is parameterized by the policy quality metric.  We consider multiple quality metrics, including number of roles, weighted structural complexity (a generalization of policy size), and (when user attribute information is available) interpretability, i.e., how well role membership can be characterized in terms of user attributes.  Ours is the first TRBAC policy mining algorithm that produces hierarchical policies, and the first that optimizes weighted structural complexity or interpretability.  In experiments with datasets based on real-world ACL policies, our algorithm is more effective than previous algorithms at optimizing policy quality.
\end{abstract}

\acmonly{
\vspace{1mm}
\noindent
{\bf Categories and Subject Descriptors:}
D.4.6 [{\bf Operating Systems}]: Security and Protection---{\it Access 
Controls};
H.2.8 [{\bf Database Management}]: Database Applications---{\it Data Mining}
}



\jcsonly{
\begin{keyword}
role mining \sep temporal role-based access control
\end{keyword}}

\section{Introduction}
\label{sec:intro}

Role-based access control (RBAC) offers significant advantages over lower-level access control policy representations, such as access control lists (ACLs).  RBAC policy mining algorithms have potential to significantly reduce the cost of migration to RBAC, by partially automating the development of an RBAC policy from an access control list (ACL) policy and possibly other information, such as user attributes \cite{hachana12role}.  The most widely studied versions of the RBAC policy mining problem involve finding a minimum-size RBAC policy consistent with (i.e., equivalent to) given ACLs.  When user attribute information is available, it is also important to maximize interpretability (or ``meaning'') of roles---in other words, to find roles whose membership can be characterized well in terms of user attributes.  Interpretability is critical in practice.  Researchers at HP Labs report ``the biggest barrier we have encountered to getting the results of role mining to be used in practice'' is that ``customers are unwilling to deploy roles that they can't understand'' \cite{ene08fast}.  Algorithms for mining meaningful roles are described in, e.g., \cite{molloy10mining,xu12algorithms}. 

Temporal RBAC (TRBAC) extends RBAC to limit the times at which roles are enabled \cite{bertino01trbac}.  TRBAC supports an expressive notation, called {\it periodic expressions}, for expressing sets of time intervals during which a role is enabled.  A role's permissions are available to members only while the role is enabled.  This allows tighter enforcement of the principle of least privilege.  Access control in many existing systems supports some form of groups or roles and some form of periodic temporal constraints.  This includes LDAP-based directory servers, such as Oracle Unified Directory and Red Hat Directory Server, XACML-based Identity and Access Management (IAM) products, such as Axiomatics Policy Server, some other IAM products, such as NetIQ Access Manager, some cloud computing services, such as Joyent's Triton Compute Service, and many network routers and switches.


This paper presents an algorithm for mining hierarchical TRBAC policies.  It is parameterized by a policy quality metric.  We consider multiple policy quality metrics: number of roles, {\it weighted structural complexity} ($\wsc$) \cite{molloy10mining}, a generalization of syntactic policy size, {\it interpretability} ($\polinterp$) \cite{molloy10mining,xu12algorithms}, described briefly above, and a compound quality metric, denoted $\polszinterp$, that combines $\wsc$ and $\polinterp$.  Our algorithm does not require attribute data; attribute data, if available, is used only in the policy quality metric, if it considers interpretability.  Our algorithm is the first TRBAC policy mining algorithm that produces hierarchical policies, and the first that optimizes WSC or interpretability.



Our algorithm is based on Xu and Stoller's elimination algorithm for RBAC mining \cite{xu12algorithms} and some aspects of Mitra {\it et al.}'s pioneering generalized temporal role mining algorithm, which we call GTRM algorithm, for mining flat TRBAC policies (i.e., policies without role hierarchy) with minimal number of roles \cite{mitra13toward,mitra15trbac}, which inspired our work.  
Our algorithm has four phases: (1) produce a set of candidate roles that contains initial roles (generated directly from the entitlements in the input) and roles created by intersecting initial roles, (2) merge candidate roles where possible, (3) organize the candidate roles into a role hierarchy, and (4) remove low-quality candidate roles\fullonly{ (this is a greedy heuristic)}.
The generated policy is not guaranteed to have optimal quality.  Fundamentally, this is because the problem of finding an optimal policy is NP-complete (this follows from NP-completeness of the untimed version of the problem (\cite{molloy10mining}).

To evaluate the algorithm, we created datasets based on real-world ACL policies from HP, described in \cite{ene08fast} and used in several evaluations of role mining algorithms, e.g., \cite{molloy10mining,xu12algorithms,mitra15trbac}.  We could simply extend the ACLs with temporal information to create a temporal user-permission assignment (TUPA), and then mine a TRBAC policy from the TUPA and attribute data.  However, it would be hard to evaluate the algorithm's effectiveness, because there is nothing with which to compare the quality of the mined policies.  Therefore, we adopt a similar methodology as Mitra {\it et al.} \cite{mitra15trbac}.  For each ACL policy, we mine an RBAC policy from the ACLs and synthetic attribute data using Xu and Stoller's elimination algorithm \cite{xu12algorithms}, pseudorandomly extend the RBAC policy with temporal information numerous times to obtain TRBAC policies, expand the TRBAC policies into equivalent TUPAs, mine a TRBAC policy from each TUPA and the attribute data, and compare the average quality of the resulting TRBAC policies with the quality of the original TRBAC policy, with the goal that the former is at least as good as the latter.

We created two datasets, using different temporal information when extending RBAC policies to obtain TRBAC policies.  For the first dataset, we use simple periodic expressions, each of which is a range of hours that implicitly repeats every day.\fullonly{ We use the same time intervals as \cite{mitra15trbac}.  They are designed to cover various relationships between intervals, such as overlapping, consecutive, disjoint, and nested.}  For the second dataset, we use more complex periodic expressions based on a hospital staffing schedule.\fullonly{ For both datasets, we use the same attribute data, namely, the high-fit synthetic attribute data for these ACL policies described in \cite{xu12algorithms}.}

In experiments using number of roles as the policy quality metric, Mitra {\it et al.}'s GTRM algorithm, designed to minimize number of roles, produces 34\% more roles than our algorithm, on average.  In experiments using $\polszinterp$ as the policy quality metric, our algorithm succeeds in finding the implicit structure in the TUPA, producing policies with comparable (for the first dataset) or moderately higher (for the second dataset) WSC and better interpretability, on average, compared with the original TRBAC policy.

Mitra {\it et al.} developed another temporal role mining algorithm, called the CO-TRAPMP-MVCL algorithm \cite{mitra16trbac}.  It minimizes a restricted variant of WSC based on the sizes of two components of the policy.  In experiments using that variant as a policy quality metric, and using datasets created by Mitra {\it et al.}, our algorithm produces policies that are 41\% smaller, on average, than the policies produced by the CO-TRAPMP-MVCL algorithm.

We explored the effect of different inheritance types on the quality of the mined policy and found that weakly restricted inheritance leads to policies with significantly better WSC and slightly better interpretability, on average.  We experimentally evaluated the benefits of some design decisions and quantified the cost-quality trade-off provided by a parameter to our algorithm that limits the set of initial roles used in intersections in phase 1.

This paper is a revised and extended version of \cite{stoller16mining}.  The main improvements are
substitution of FastMiner for CompleteMiner when computing role intersections and an empirical justification for this, an improved metric for selecting a subset of initial roles for use in role intersections, more explanation and details of the algorithm, and more experiments, including an experimental comparison with Mitra {\it et al.}'s CO-TRAPMP-MVCL algorithm \cite{mitra16trbac}.

Section \ref{sec:trbac} provides background on TRBAC.  Section \ref{sec:problem} defines the policy mining problem.  Section \ref{sec:algorithm} presents our algorithm.  Section \ref{sec:datasets} describes the datasets used in the experimental evaluation.  Section \ref{sec:eval} presents the results of the experimental evaluation.  Section \ref{sec:related} discusses related work.  Directions for future work include: mining TRBAC policies from operation logs, by extending work on mining RBAC policies from logs \cite{molloy12generative}; optimization of TRBAC policies, i.e., improving the quality of a TRBAC policy while minizing changes to it, by extending work on optimizing RBAC policies \cite{vaidya08migrating}; and mining temporal ABAC policies, by extending work on ABAC policy mining \cite{xu15miningABAC,medvet2015}.


\section{Background on TRBAC}
\label{sec:trbac}

An {\em RBAC policy} is a tuple $\tuple{\User, \Perm, \Role, \ua, \pa, \rh}$, where $\User$ is a set of users, $\Perm$ is a set of permissions, $\Role$ is a set of roles, $\ua\subseteq \User \times \Role$ is the user-role assignment, $\pa\subseteq \Role \times \Perm$ is the permission-role assignment,
and $\rh \subseteq \Role\times \Role$ is the role inheritance relation (also called
the role hierarchy).  Specifically, $\tuple{r,r'}\in \rh$ means that $r$ is
senior to $r'$, hence all permissions of $r'$ are also permissions of $r$,
and all members of $r$ are also members of $r'$.
A role $r'$ is {\em junior to} role $r$ if $r \rh^+ r'$, where $\rh^+$ is
the transitive closure of $\rh$.

A {\em periodic expression} (PE) is a symbolic representation for an infinite set of time intervals.  The formal definition of periodic expressions in \cite{bertino01trbac,mitra15trbac} is standard and somewhat complicated; instead of repeating it, we give a brief intuitive version.  A {\em calendar} is an infinite set of consecutive time intervals of the same duration; informally, it corresponds to a time unit, e.g., a day or an hour.  A sequence of calendars $C_1,\ldots,C_n,C_d$ defines the sequence of time units used in a periodic expression, from larger to smaller.  A periodic expression has the form $\sum_{k=1}^n O_k \cdot C_k \;\rhd\; d \cdot C_d$ where $O_1=\all$, $O_k$ is a set of natural numbers or the special value $\all$ for $2 \le k \le n$, and $d$ is a natural number.  The first part of a PE (before $\rhd$) identifies the set of starting points of the intervals represented by the PE.  The second part of the PE (after $\rhd$) specifies the duration of each interval.

For example, consider the sequence of calendars Quadweeks, Weeks, Days, hours, where a Quadweek is four consecutive weeks---similar to a month, but with a uniform duration.  The periodic expression [$\all$ $\cdot$ Quadweeks + \{1,3\} $\cdot$ Weeks + \{1,2,3,4,5\} $\cdot$ Days + \{10\} $\cdot$ Hours $\rhd$ 8 $\cdot$ Hours] represents the set of time intervals starting at 9am (the time intervals in each calendar are indexed starting with 1, so for Hours, 1 denotes the hour starting at midnight, 2 denotes the hour starting at 1am, etc.) and ending at 5pm (since duration is 8 hours) of every weekday (assuming days of the week are indexed with 1=Monday) during the first and third weeks of every quadweek.

A {\em bounded periodic expression} (BPE) is a tuple $\tuple{[{\it begin}, {\it end}], \pe}$, where {\it begin} and {\it end} are date-times, and $\pe$ is a periodic expression.  A BPE represents the set of time intervals represented by $pe$ except limited to the interval $[{\it begin}, {\it end}]$.

A {\em BPE set} (BPES) is a set of BPEs.  It represents the union of the sets of time intervals represented by its members

A {\em temporal RBAC (TRBAC) policy} is a tuple $\langle \User, \Perm, \Role, \ua, \pa, \rh,$ $\inhty, {\it REB}\rangle$, where the first six components are the same as for an RBAC policy, $\inhty$ is the inheritance type (described below), and ${\it REB}$ is the role enabling base (REB), which specifies when roles are enabled \cite{bertino01trbac}.  Bertino {\it et al.} allow the REB to specify various conditions and events that enabled or disable a role.  Like Mitra {\it et al.} \cite{mitra15trbac,mitra16trbac}, we are interested only in temporal conditions and therefore consider a limited form of REB, which we call a {\it role-time assignment}.  Specifically, a role-time assignment $\ta$ maps each role to a BPES.  A role $r$ is enabled during the set of time intervals represented by $\ta(r)$.  A REB can easily be constructed from a role-time assignment, so an RBAC policy with temporal conditions represented by a role-time assignment instead of a REB can also be considered a TRBAC policy.


We consider two types of inheritance \cite{joshi2002temporal}.  In both cases, a senior role $r$ inherits permissions from each of its junior roles $r'$.  With {\it weakly restricted inheritance}, denoted by $\inhty={\rm WR}$, a permission inherited from $r'$ is available to members of $r$ during the time intervals specified by $\ta(r)$.  With {\it strongly restricted inheritance}, denoted by $\inhty={\rm SR}$, a permission inherited from $r'$ is available to members of $r$ during the time intervals specified by $\ta(r')$.


A {\em temporal user-permission assignment (TUPA)} is a set of triples of the form $\tuple{u, p, \bpes}$, where $u$ is a user, $p$ is a permission, and $\bpes$ is a BPES.  We refer to such a triple as an {\it entitlement triple}.  Such a triple means that $u$ has permission $p$ during the set of time intervals represented by $\bpes$.  A TUPA should contain at most one entitlement triple for each user-permission pair.  A TUPA can therefore be regarded as a dictionary that maps user-permission pairs to BPESs.

The meaning of a role $r$ in a TRBAC policy $\pi$, denoted $\mean{r}_\pi$, is a TUPA that expresses the entitlements granted by $r$, taking inheritance into account.  The meaning $\mean{\pi}$ of a TRBAC policy $\pi$ is a TUPA that expresses the entitlements granted by $\pi$.

%



\section{The Relaxed TRBAC Policy Mining Problem}
\label{sec:problem}

A {\it policy quality metric} is a function from TRBAC policies to a totally-ordered set, such as the natural numbers.  The ordering is chosen so that small values indicate high quality\fullonly{; this might seem counter-intuitive at first glance, but it is natural for metrics such as policy size}.\fullonly{ We define three basic policy quality metrics and then consider combinations of them.}

{\it Number of roles} is a simplistic but traditional policy quality metric.

{\it Weighted Structural Complexity} (WSC) is a generalization of policy size \cite{molloy10mining}.  For a TRBAC policy $\pi$ of the above form with a role-time assignment $\ta$ as its REB, we define the WSC of $\pi$ to be $\wsc(\pi) = w_1|\Role| + w_2|\ua| + w_3|\pa| + w_4|\rh| + w_5\wsc(\ta)$, where the $w_i$ are user-specified weights, $|s|$ is the size (cardinality) of set $s$, and $\wsc(\ta)$ is the sum of the sizes of the BPESs in $\ta$.  The size of a BPES is the sum of the sizes of the BPEs in it.  The size of a BPE is the size of the PE in it (the beginning and ending date-times have fixed size, so we ignore them).  The size of a PE is the sum of the sizes of the sets in it plus 1 for the duration, with the special value $\all$ counted as a set of size 1.


{\it Interpretability} is a policy quality metric that measures how well role membership can be characterized in terms of user attributes.  {\it User-attribute data} is a tuple $\tuple{A,f}$, where $A$ is a set of attributes, and $f$ is a function such that $f(u,a)$ is the value of attribute $a$ for user $u$.
An {\it attribute expression} $e$ is a function from the set $A$ of attributes to sets of values.  A user $u$ {\it satisfies} an attribute expression $e$ iff $(\forall a \in A.\; f(u,a) \in e(a))$.  For example, if $A=\{{\it dept}, {\it level}\}$, the function $e$ with $e({\it dept})=\{{\rm CS}\}$ and $e({\it level})=\{2,3\}$ is an attribute expression, which can be written with syntactic sugar as ${\it dept}\in\{{\rm CS}\} \,\land\, {\it level}\in\{2,3\}$.  We refer to the set $e(a)$ as the conjunct for attribute $a$.  Let $\su{e}$ denote the set of users that satisfy $e$.  For an attribute expression $e$ and a set $U$ of users, the {\it mismatch} of $e$ and $U$ is defined by ${\rm mismatch}(e, U) = |\su{e} \ominus U|$, where the symmetric difference of sets $s_1$ and $s_2$ is $s_1 \ominus s_2 = (s_1 \mysetminus s_2)\union (s_2\mysetminus s_1)$.  The {\it attribute mismatch} of a role $r$, denoted $\am(r)$, is $\min_{e\in E} {\rm mismatch}(e, \assignedu(r))$, where $E$ is the set of all attribute expressions, and $\assignedu(r)=\{u \;|\; \tuple{u,r}\in\ua\}$.  An attribute expression $e$ that minimizes the attribute mismatch of role $r$ is called a {\em best-fit attribute expression for $r$}.  Intuitively, it is the most accurate possible ``explanation'' (characterization) of $r$'s membership using the given attribute data; it can be shown to users to help them understand the role.  We define policy interpretability $\polinterp$ as the sum over roles of attribute mismatch, i.e., $\polinterp(\pi) = \sum_{r \in \Role} \am(r)$.

{\it Compound policy quality metrics} take multiple aspects of policy quality into account.  We combine metrics by Cartesian product, with lexicographic order on the tuples.  Lexicographic order means $\tuple{x_1,y_1} < \tuple{x_1,y}$ iff $x_1 < x_2$ or $x_1=x_2 \land y_1 < y_2$.  Weighted sums of policy quality metrics could also be used.  Let
$\polszinterp(\pi)=\tuple{\wsc(\pi), \polinterp(\pi)}$.


A TRBAC policy $\pi$ is {\em consistent} with a TUPA $T$ if they grant the same permissions to the same users for the same sets of time intervals.  When the given TUPA contains noise, it is desirable to weaken this requirement.  A TRBAC policy $\pi$ is {\em $\epsilon$-consistent} with a TUPA $T$, where $\epsilon$ is a natural number, if they grant the same permissions to the same users for the same sets of time intervals, except that, for at most $\epsilon$ entitlement triples $\tuple{u,p,\bpes}$ in $T$, the policy $\pi$ either does not grant $p$ to $u$ or grants $p$ to $u$ at fewer times than $\bpes$ \cite{mitra15trbac}.  Note that consistency is a special case of $\epsilon$-consistency, corresponding to $\epsilon=0$.

The {\it relaxed TRBAC policy mining problem} is: given a TUPA $T$, policy quality metric $\qpol$, and consistency threshold $\epsilon$, find a TRBAC policy $\pi$ that is $\epsilon$-consistent with $T$ and has the best quality, according to $\qpol$, among policies $\epsilon$-consistent with $T$.   Auxiliary information used by the policy quality metric, e.g., user-attribute data, is implicitly considered to be part of $\qpol$ in this definition.  Note that the temporal part of $T$ strongly influences $\pi$, even using WSC with $w_5=0$, because it determines how entitlements can be grouped in roles.

\fullonly{We refer to this as the {\em relaxed} TRBAC policy mining problem, because of the relaxed consistency requirement; Mitra {\it et al.} refer to it as the {\em generalized} TRBAC policy mining problem.}

\myparagraph{Suggested role assignments for new users}

If attribute data is available, the system can compute and store a best-fit attribute expression $e_r$ for each role $r$.  When a new user $u$ is added, the system can suggest that $u$ be made a member of the roles for which $u$ satisfies the best-fit attribute expression, and it presents these suggested roles in ascending order of attribute mismatch.  This reduces the administrative effort involved in assigning roles to new users.


\section{TRBAC Policy Mining Algorithm}
\label{sec:algorithm}

Inputs to the algorithm are the TUPA $T$, the type of inheritance $\inhty$ to use in the generated policy, the consistency threshold $\epsilon$, and the policy quality metric $\qpol$.  \fullonly{ While reading the TUPA, our algorithm attempts to simplify the BPES in each triple by merging BPEs in it that represent sets of overlapping or consecutive time intervals; this is done in the same way as in case (2b) of Phase 2, described below.}

In traditional RBAC and TRBAC notation, roles are identifiers (not objects), and separate relations such as $\ua$ (not object attributes) provide information about them.  Similarly, in our pseudocode, roles have no attributes; instead, dictionaries map roles to relevant information.

Our pseudocode uses the following notation for sets and dictionaries.  ``new Set()'' and ``new Dictionary()'' create an empty set and empty dictionary, respectively.  The methods of a set $s$ include $s.{\rm add}(x)$ to add an element $x$, $s.{\rm remove}(x)$ to remove an element $x$, $s.{\rm addAll}(x)$ to add all elements of set $s_2$, and $s.{\rm copy}(x)$ to create a copy of $x$.  The statement $d(k)=v$ updates dictionary $d$ to map key $k$ to value $v$.  The expression $d(k)$ returns the value that dictionary $d$ associates with key $k$; it is used only in contexts where $d$ contains an entry for $k$.

\myparagraph{Phase 1: Generate roles}

Phase 1 generates initial roles and then creates additional candidate roles
by intersecting sets of initial roles.

\myparagraph{Phase 1.1: Generate initial roles}

Pseudocode for generating initial roles appears in Figure \ref{fig:generate-init}.  It uses a semantic containment relation $\sqsubseteq$ on PEs, BPEs, and BPESs: $x_1 \sqsubseteq x_2$ iff the set of time instants represented by $x_1$ is a subset of the set of time instants represented by $x_2$.  Note that, for BPESs $\bpes_1$ and $\bpes_2$, $\bpes_1 \sqsubseteq \bpes_2$ may hold even if $\bpes_1 \subseteq \bpes_2$ does not hold.  The function permBPES groups together the set of permissions $P$ that a user $u$ has for exactly the same BPES $\bpes$ or a BPES $\bpes'$ that semantically contains $\bpes$. An initial role is created with user $u$, the resulting set of permissions $P$, and time assignment $\bpes$. In addition, for each BPE $\bpe$ in $\bpes$, we create an initial role with user $u$, permissions $P$, and time assignment $\set{\bpe}$.

\begin{figure}[tb]
\centering
\fbox{\begin{minipage}[t]{\linewidth}
\begin{tabbing}
$\rinit$ = new Set()\\
$\assignedpz$ = new Dictionary()\\
$\assigneduz$ = new Dictionary()\\
$\taz =$ new Dictionary()\\
\forloop~$u$ in $U$\\
~~\= \forloop~$\tuple{P, \bpes}$ in permBPES($u, T$)\\
    \> ~~\= addRole($\rinit,\set{u}, P, \bpes$)\\
\>\>        \forloop~$\bpe$ in $\bpes$\\
\>\> ~~\=     addRole($\rinit,\set{u}, P, \set{\bpe}$)\\
\\
permBPES($u, T$) =\\
 $\{ \tuple{P,\bpes} \;|\; $
  \= $(\exists p. \tuple{u, p, \bpes} \in T)$\\
  \> ${}\land P = \{p \;|\; $ \= $\tuple{u, p, \bpes'} \in T$\\
\>\> $\land \bpes \sqsubseteq \bpes'\}\}$\\
\end{tabbing}
\end{minipage}}
\fbox{\begin{minipage}[t]{\linewidth}
\begin{tabbing}
\function~addRole($R,U,P,\bpes$)\\
\com{if there is an existing role with permissions $P$ and}\\
\com{BPES $\bpes$, add users in $U$ to it, else create new}\\
\com{role with users $U$, permissions $P$, and BPES $\bpes$.}\\
\ifstmt~$U$, $P$, or $\bpes$ is empty\\
~~ \return\\
\ifstmt~$\exists$ $r$ in $R$ s.t. \=$\assignedpz(r)=P \land \taz(r)=\bpes$\\
~~ $\assigneduz(r)$.addAll($U$)\\
\elsestmt\\
~~ \= $r$ = new Role()\\
\> $\assignedpz(r) = P$\\
\> $\assigneduz(r) = U$\\
\> $\taz(r) = \bpes$\\
\> $R$.add($r$)
\end{tabbing}
\end{minipage}}
\vspace{-1ex}
\caption{Phase 1.1: Generate initial roles.  ``s.t.'' abbreviates ``such
  that''.}
\label{fig:generate-init}
\end{figure}

\myparagraph{Phase 1.2: Intersect roles}

Phase 1.2 starts to construct a set $\rcand$ of candidate roles, by adding to $\rcand$ all of the initial roles in $\rinit$ and all non-empty intersections of all pairs of initial roles.  In other words, for each pair of initial roles, if the intersection of their permission sets is a non-empty set $P$, and the intersection of their BPESs is a non-empty BPES $\bpes$, then create a candidate role with permissions $P$, BPES $\bpes$, and the union of their user sets.  BPESs are intersected semantically, not syntactically; for example, if $\bpes_1$ represents 9am-5pm on Mondays and Wednesdays, and $\bpes_2$ represents 1pm-2pm on Mondays and Fridays, then their intersection is a BPES that represents 1pm-2pm on Mondays.  This phase is similar to role intersection in FastMiner \cite{vaidya06roleminer}. Pseudocode appears in Figure \ref{fig:role-intersection}.  The function $\sqcap$ denotes semantic intersection of BPESs; in other words, $\bpes_1 \sqcap \bpes_2$ is a BPES that represents the set of time instants represented by $\bpes_1$ and $\bpes_2$.

\begin{figure}[tb]
\fbox{\begin{minipage}[t]{\linewidth}
\begin{tabbing}
\forloop~r in $\rinit$\\
~~\= $\rinit$.remove($r$)\\
\> $\rcand$.add($r$)\\
\> \forloop~$r'$ in $\rinit$\\
\> ~~\= $P = \assignedpz(r) \intersect \assignedpz(r')$\\
\>\> $\bpes = \taz(r) \sqcap \taz(r')$\\
\>\> \ifstmt~$P$ and $\bpes$ are non-empty\\
\>\>~~ addRole($\rcand,\assigneduz(r) \union \assigneduz(r'), P, \bpes$)
\end{tabbing}
\end{minipage}}
\vspace{-1ex}
\caption{Phase 1.2: Intersect roles}
\label{fig:role-intersection}
\end{figure}

\comment{
\begin{figure}[tb]
\fbox{\begin{minipage}[t]{\linewidth}
\begin{tabbing}
$\rcand$ = $\rinit$.copy()\\
$\rvis$ = new Set()\\
\forloop~r in $\rinit$\\
~~\= $\rvis$.add($r$)\\
\> \forloop~$r'$ in $\rcand\setminus\rvis$\\
\> ~~\= $P = \assignedpz(r) \intersect \assignedpz(r')$\\
\>\> $\bpes = \taz(r) \sqcap \taz(r')$\\
\>\> \ifstmt~$P$ and $\bpes$ are non-empty\\
\>\>~~ addRole($\rcand,\assigneduz(r) \union \assigneduz(r'), P, \bpes$)
\end{tabbing}
\end{minipage}}
\vspace{-1ex}
\caption{Phase 1.2: Intersect roles}
\label{fig:role-intersection}
\end{figure}}

This phase is expensive for large datasets.  To reduce the cost, we allow role intersections to be limited to a subset of the initial roles containing the roles mostly likely to produce useful intersections.  To support a flexible trade-off between cost (running time) and policy quality, we introduce a parameter that controls the size of the subset. 

The subset is characterized using a new role quality metric, called the {\em usefulness-for-intersection metric} ($\ui$ metric).  It is a weighted sum of four quantities relevant to the usefulness of a role $r$ in intersections: role size (sum of number of users, number of permissions, and the WSC of the BPES), $\coveredup(r)$ (defined below), permission popularity (sum over the permissions $p$ of $r$ of the fraction of initial roles having permission $p$), and PE popularity (sum over the PEs $\pe$ in $r$'s BPES of the fraction of initial roles having $\pe$ in its BPES).  For example, consider the set of roles $\set{r_1, r_2, r_3}$, where $r_1$ has permissions $\set{p_1, p_2}$ and enabled time $\set{\pe_1}$, $r_2$ has permissions $\set{p_1}$ and enabled time $\set{\pe_1}$, and $r_3$ has permissions $\set{p_4}$ and enabled time $\set{\pe_2,\pe_3}$ (user assignments are irrelevant hence omitted).  The permission popularity of $r_1$ is ${2\over 3} + {1\over 3} = 1$, of $r_2$ is ${2\over 3}$, and of $r_3$ is ${1\over 3}$.  The PE popularity of $r_1$ is ${2\over 3}$, of $r_2$ is ${2\over 3}$, and of $r_3$ is ${1\over 3} + {1\over 3} = {2\over 3}$.

We used a Support Vector Machine (SVM) to find the weights that maximize the $\ui$ metric's effectiveness as a classifier for whether an initial role is ``useful for intersections'', i.e., is used in an intersection that contributes to the final policy, either directly or via merges.  We extended our system to keep track of which initial roles are useful for intersections, ran the extended system on one small policy (domino), and trained the SVM on the resulting data.  The resulting weights are -2.7357, -1.6484, 2.3417, and -0.6017, respectively.  The signs of the parameters show that, for example, roles with smaller size and more popular permissions are more useful in intersections.



To control the cost-quality trade-off, we introduce a parameter $\ric$ (mnemonic for ``role intersection cutoff'') that ranges between 0 and 1, sort the roles by the usefulness-for-intersection metric, and use only roles in the top $\ric$ in intersections.  For example, $\ric=0.3$ means that only roles whose values of the $\ui$ metric are in the top (\ie, largest) 30\% are used in intersections.





\myparagraph{Phase 2: Merge roles}

Phase 2 merges candidate roles to produce a revised set of candidate roles.
It uses the following types of merges.
(1) If candidate roles $r$ and $r'$ have the same set of users $U$ and the
same BPES $\bpes$, then they are replaced with a new role with users $U$,
permissions $\assignedpz(r)\union\assignedpz(r')$, and BPES $\bpes$\fullonly{, unless a role with those permissions and that BPES already exists, in which case the users $U$ are added to it}.
(2) If candidate roles $r$ and $r'$ have the same users $U$ and same permissions $P$, then they are replaced with a new role with users $U$, permissions $P$, and BPES $\bpes(r) \sqcup \bpes(r')$\fullonly{, unless a role with those permissions and that BPES already exists, in which case the users $U$ are added to it}.\fullonly{ Pseudocode appears in Figure \ref{fig:merge}.} The function $\sqcup$ denotes semantic union of BPESs; in other words, $\bpes_1 \sqcup \bpes_2$ is a BPES that represents the set of time instants represented by $\bpes_1$ or $\bpes_2$.
We distinguish two sub-cases.
(2a) If $\bpes_1$ and $\bpes_2$ represent disjoint sets of time intervals,
then $\bpes_1 \sqcup \bpes_2$ is simply $\bpes_1 \union \bpes_2$.
(2b) If $\bpes_1$ and $\bpes_2$ represent sets of overlapping or consecutive
time intervals, then BPEs in them are merged, if possible, to simplify
the result.  For example, if $\bpes_1$ represents 9am-noon on weekdays, and
$\bpes_2$ denotes noon-5pm on weekdays, then $\bpes_1 \sqcup \bpes_2$ contains
a single BPE denoting 9am-5pm on weekdays.

\fullonly{
\begin{figure}[tb]
\centering
\fbox{\begin{minipage}[t]{\linewidth}
\begin{tabbing}
$\rvis$ = new Set()\\
\forloop~r in $\rcand$\\
~~\= $\rvis$.add($r$)\\
\> \forloop~$r'$ in $\rcand\setminus\rvis$\\
\>~~\= mergeIfSameMemberBPES($\rcand,r,r'$)\\
\>\> mergeIfSameMemberPerm($\rcand,r,r'$)\\
\\
\function~mergeIfSameMemberBPES($\rcand,r,r'$)\\
   \ifstmt~$
\begin{array}[t]{@{}l@{}}
  \assigneduz(r) = \assigneduz(r')\\
  {}\land \ta(r) = \ta(r')
\end{array}$\\
~~\=   \ifstmt~$\assignedpz(r) \subseteq \assignedpz(r')$\\
\>~~\=     \com{merging $r$ and $r'$ yields $r'$, so just remove $r$}\\
\>\>       $\rcand.{\rm remove}(r)$\\
\>      \elsestmt~\ifstmt~$\assignedpz(r') \subseteq \assignedpz(r)$\\
\>\>       \com{merging $r$ and $r'$ yields $r$, so just remove $r'$}\\
\>\>       $\rcand.{\rm remove}(r')$\\
\>      \elsestmt\\
\>\>       $P = \assignedpz(r) \union \assignedpz(r')$\\
\>\>	   \ifstmt~$\exists$ $r''$ in $\rcand$ s.t. $\assignedpz(r'')=P$\\
\>\>         ~~~~~~~~~~~~~~~~~${} \land \ta(r'')=\ta(r)
$\\
\>\>~~\=	    $\assigneduz(r'').{\rm addAll}(\assigneduz(r))$\\
\>\>	   \elsestmt\\
\>\>\>          $r''$ = new Role()\\
\>\>\>          $\assigneduz(r'') = \assigneduz(r)$\\
\>\>\>          $\assignedpz(r'') = P$\\
\>\>\>          $\ta(r'') = \ta(r)$\\
\>\>\>          $\rcand.{\rm add}(r)$\\
\>\>       $\rcand.{\rm remove}(r)$\\
\>\>       $\rcand.{\rm remove}(r')$
\end{tabbing}
\end{minipage}}
\fbox{\begin{minipage}[t]{\linewidth}
\begin{tabbing}
\function~mergeIfSameMemberPerm($\rcand,r,r'$)\\
		\ifstmt~$
\begin{array}[t]{@{}l@{}}
  \assigneduz(r) = \assigneduz(r')\\
  {} \land \assignedpz(r) = \assignedpz(r')
\end{array}$\\
~~\=        \ifstmt~$\ta(r) \sqsubseteq \ta(r')$\\
\>~~\=           $\rcand.{\rm remove}(r)$\\
\>          \elsestmt~\ifstmt~$\ta(r') \sqsubseteq \ta(r)$\\
\>\>            $\rcand.{\rm remove}(r')$\\
\>          \elsestmt\\
\>\>			$\bpes = \ta(r) \sqcup \ta(r')$\\
\>\>            \ifstmt~$\exists$ $r''$ in $R$ s.t. $
\begin{array}[t]{@{}l@{}}
  \assignedpz(r'')=\assignedpz(r)\\ 
  {} \land \ta(r'')=\bpes
\end{array}$\\
\>\>~~\=        $\assigneduz(r'').{\rm addAll}(\assigneduz(r))$\\
\>\>	   \elsestmt\\
\>\>\>          $r''$ = new Role()\\
\>\>\>          $\assigneduz(r'') = \assigneduz(r)$\\
\>\>\>          $\assignedpz(r'') = \assignedpz(r)$\\
\>\>\>          $\ta(r'') = \bpes$\\
\>\>\>          $\rcand.{\rm add}(r)$\\
\>\>       $\rcand.{\rm remove}(r)$\\
\>\>       $\rcand.{\rm remove}(r')$
\end{tabbing}
\end{minipage}}
\vspace{-1ex}
\caption{Phase 2: Merge roles.}
\label{fig:merge}
\end{figure}}

\myparagraph{Phase 3: Construct role hierarchy}

Phase 3 organizes the candidate roles into a role hierarchy with full inheritance.  A TRBAC policy has {\it full inheritance} if every two roles that can be related by the inheritance relation are related by it, i.e., $\forall r, r' \in R.\; \mean{r}_{\pi} \supseteq \mean{r'}_\pi \myimplies \tuple{r,r'}\in\rh^*$.  Guo {\it et al.} call this property {\it completeness} in the context of RBAC \cite{guo08role}.\fullonly{ We always generate policies with full inheritance, even though relaxing this requirement would allow our algorithms to achieve better policy quality in some cases, because in the absence of other information, all of these possible inheritance relationships are equally plausible, and removing any of them risks removing some that are semantically meaningful and desirable.}

\myparagraph{Phase 3.1: Compute inheritance}

Phase 3.1 determines inheritance relationships between candidate roles, based on the requirement of full inheritance.  Function $\isancestorfi(r',r)$ tests whether $r'$ is an ancestor of $r$ with full inheritance; if $\inhty={\rm WR}$, the function avoids inheritance relationships that would lead to cycles in the role hierarchy.
\begin{displaymath}
\begin{array}[t]{@{}l@{}}
\isancestorfi(r',r) = \\
~~  \assignedpz(r') \subseteq \assignedpz(r) \land \assigneduz(r) \subseteq \assigneduz(r')\\
~~ {} \land (\inhty={\rm SR} \myimplies \taz(r') \sqsubseteq \taz(r))\\
~~ {} \land (\inhty={\rm WR} \myimplies \neg (\assignedpz(r) \subset \assignedpz(r') 
             \land \assigneduz(r') \subset \assigneduz(r)))  
\end{array}
\end{displaymath}
This function is called for every pair of candidate roles.  If $\isancestorfi(r',r)$ is true, and there is no role between $r'$ and $r$ in the role hierarchy (i.e., no role $r''$ such that $\isancestorfi(r', r'')$ and $\isancestorfi(r'', r)$), then $r'$ is a parent of $r$.  This phase produces dictionaries $\parents$ and $\children$, such that $\parents(r)$ and $\children(r)$ are the sets of parents and children of $r$, respectively.  \fullonly{ Pseudocode appears in Figure \ref{fig:create-inher}.}

\fullonly{
\begin{figure}[tb]
\fbox{\begin{minipage}[t]{\linewidth}
\begin{tabbing}
$\parents = $new Dictionary()\\
$\children = $new Dictionary()\\
\forloop~$r$ in $\rcand$\\
~~ \parents($r$) = new Set()\\
~~ \children($r$) = new Set()\\
\forloop~$r$ in $\rcand$\\
~~\= \forloop~$r'$ in $\rcand\setminus\set{r}$\\
\> ~~\= \ifstmt~\isancestorfi($r', r$)\\
\>\>~~\= \com{check whether $r'$ is a parent or a more distant ancestor of $r$}\\
\>\>\> \ifstmt~$\neg\exists$ $r''$ in $\parents(r)$ s.t. $\isancestorfi(r',r'')$\\
\>\>\>~~\= \com{$r'$ is a parent of $r$, based on roles considered so far.}\\
\>\>\>\>\com{a subsequent role could be placed between them.}\\
\>\>\>\> \parents($r$).add($r'$)\\
\>\>\>\> \com{remove parents of $r$ that are also parents of $r'$.}\\
\>\>\>\> \forloop~$r''$ in $\parents(r)\setminus\set{r'}$\\
\>\>\>\> ~~\= \ifstmt~\isancestorfi($r'',r')$\\
\>\>\>\>\> \parents($r$).remove($r''$)\\
\>\> \ifstmt~\isancestorfi($r, r'$)\\
\>\>~~\= \com{check whether $r'$ is a child or more distant descendant of $r$}\\
\>\>\> \ifstmt~$\neg\exists$ $r''$ in $\children(r)$ s.t. $\isancestorfi(r'',r')$\\
\>\>\>~~\= \com{$r'$ is a child of $r$, based on roles considered so far.}\\
\>\>\>\>\com{a subsequent role could be placed between them.}\\
\>\>\>\> \children($r$).add($r'$)\\
\>\>\>\> \com{remove children of $r$ that are also children of $r'$.}\\
\>\>\>\> \forloop~$r''$ in $\children(r)\setminus\set{r'}$\\
\>\>\>\> ~~\= \ifstmt~\isancestorfi($r',r'')$\\
\>\>\>\>\> \children($r$).remove($r''$)
\end{tabbing}
\end{minipage}}
\vspace{-1ex}
\caption{Phase 3.1: Determine inheritance relationships.}
\label{fig:create-inher}
\end{figure}}

\myparagraph{Phase 3.2: Compute assigned users and permissions}

Phase 3.2 computes the directly assigned users $\assignedu(r)$ and directly assigned permissions $\assignedp(r)$ of each role $r$, by removing inherited users and permissions from the role's originally assigned users $\assigneduz(r)$ and originally assigned permissions $\assignedpz(r)$.\fullonly{ Pseudocode appears in Figure \ref{fig:directly-assigned-UP}.}

\fullonly{\begin{figure}[tb]
\fbox{\begin{minipage}[t]{\linewidth}
\begin{tabbing}
\forloop~$r$ in $\rcand$\\
~~\= {\it inheritedU} = $\UNION_{r' in \children(r)} \assigneduz(r')$\\
\> \assignedu($r$) = $\assigneduz(r)$.copy().removeAll({\it inheritedU})\\
\> \ifstmt~\inhty=WR\\
\>~~\=  {\it inheritedP} = $\UNION_{r' in \parents(r)} \assignedpz(r')$\\
\>\> $\assignedp(r)$ = $\assignedpz(r)$.copy().removeAll({\it inheritedP})\\
\>\ifstmt~\inhty=SR\\
\>~~\= \assignedp($r$) = $\assignedpz(r)$.copy()\\
\>\> \forloop~$p$ in $\assignedpz(r)$\\
\>\> ~~\= \com{inherBPES is the BPES with which $p$ is inherited by $r$}\\
\>\>\> {\it inherBPES} = $\sqcup_{r' \in \parents(r)} \taz(r')$\\
\>\>\> \com{if {\it inherBPES} equals $\ta(r)$, then $p$ does not need to be directly assigned, i.e., $p$ is inherited.}\\
\>\>\> \ifstmt~$\ta(r) = {\it inherBPES}$\\
\>\>\> ~~ \assignedp.remove($p$)
\end{tabbing}
\end{minipage}}
\vspace{-1ex}
\caption{Phase 3.2: Compute directly assigned users and directly assigned
  permissions.}
\label{fig:directly-assigned-UP}
\end{figure}}

\myparagraph{Phase 4: Remove roles}

Phase 4 removes roles from the candidate role hierarchy if the removal
preserves $\epsilon$-consistency with the given ACL policy and improves policy
quality.  When a role $r$ is removed, the role hierarchy is adjusted to
preserve inheritance relations between parents and children of $r$, and the
sets of directly assigned users and permissions of other roles are expanded
to contain users and permissions that they previously inherited from $r$.

The order in which roles are considered for removal affects the final result.  We control this ordering with a {\it role quality metric} $\qrole$, which maps roles to an ordered set, with the interpretation that large values denote high quality (note: this is opposite to the interpretation of the ordering for policy quality metrics).  Low-quality roles are considered for removal first.  We use a role quality metric that is a temporal variant of the role quality metric in \cite{xu12algorithms} that gave the best results in their experiments.  We define some auxiliary functions then role quality.

The {\em redundancy} of a role $r$ measures how many other roles also cover the entitlement triples covered by $r$.  We say that a role $r$ {\em covers} an entitlement triple $t$ if $t \in \mean{r}_\pi$.
Removing a role with higher redundancy is less likely to prevent subsequent removal of other roles, so we eliminate roles with higher redundancy first.  The redundancy of role $r$, denoted $\redun(r)$, is the negative of the minimum, over entitlement triples $\tuple{u, p, \bpes}$ covered by $r$, of the number of removable roles that cover $\tuple{u, p, \bpes}$ (we take the negative so that roles with more redundancy have lower quality).  A role is {\em removable} in policy $\pi$, denoted $\removable(r)$ (the policy is an implicit argument), if the policy obtained by removing $r$ is $\epsilon$-consistent with $T$.\fullonly{
\begin{eqnarray*}
  \label{eq:redun}
  \redun(\tuple{u, p, \bpes}) &=& 
    |\{r \in \rcand \;|\; \tuple{u, p, \bpes'} \in \mean{r}_\pi 
    ~{}\land \bpes \sqsubseteq \bpes' \land \removable(r)\}|\\
  \redun(r) &=& - \min_{t \in \mean{r}_\pi } (redun(t))  
\end{eqnarray*}}

The {\it clustered size} of a role $r$ measures how many entitlements are covered by $r$ and how well they are clustered.  A first attempt at formulating this metric (ignoring clustering) might be as the fraction of entitlement triples in $T$ that are covered by $r$.  As discussed in \cite{xu12algorithms}, it is better for the covered entitlement triples to be ``clustered'' on (i.e., associated with) fewer users rather than being spread across many users.  The clustered size of $r$ is defined to equal the fraction of the entitlements of $r$'s members that are covered by $r$.  In the temporal case, each entitlement triple $\tuple{u, p, \bpes}$ is weighted by the fraction of the time represented $\bpes$ that is covered by $\ta(r)$.

\begin{displaymath}
\coveredup(r) = \!\!
\sum_{
  \begin{array}{@{}l@{}}
    \scriptstyle
    u \in \assignedu(r)\vspace{-0.8ex}\\
    \scriptstyle
    p \in \assignedp(r)
  \end{array}} 
\!\!{\duration(\ta(r)) \over \duration(T(u,p))}
~~~~~~~~
\clussz(r) = {\coveredup(r) \over |\entit(\assignedu(r), T)|}
\end{displaymath}
where $T(u,p)$ is the BPES $\bpes$ such that $\tuple{u, p, \bpes}\in T$, 
$\duration(\bpes)$ is the fraction of one time unit in calendar $C_1$ that
is covered by $\bpes$, and $\entit(U,T)$ is the set of entitlement triples in $T$ for a user in $U$.  For example, if the sequence of calendars is
$C_1={\rm Year}, \ldots, C_n={\rm Hour}, C_d={\rm Hour}$, and $\bpes$ is
9am-5pm every day, then $\duration(\bpes) = 1/3$, since $\bpes$ covers 1/3 of
the time in a year.

Our role quality metric is $\qrole(r)=\tuple{\redun(r),\clussz(r)}$, with lexicographic order on the tuples.

Our algorithm may remove a role even if the removal worsens policy quality
slightly.  Specifically, we introduce a {\it quality change tolerance}
$\delta$, with $\delta\ge 1$, and we remove a role if the quality $Q'$ of
the TRBAC policy resulting from the removal is related to the quality $Q$ of
the current TRBAC policy by $Q' < \delta Q$ (recall that, for policy quality
metrics, smaller values are better).
Choosing $\delta > 1$ partially compensates for the fact that a purely
greedy approach to policy quality improvement is not an optimal strategy.

Pseudocode for removing roles appears in Figure \ref{fig:remove-roles}.  It repeatedly tries to remove all removable roles, until none of the attempted removals succeeds in improving the policy quality.  The policy $\pi$ is an implicit argument to auxiliary functions such as $\removerole$ and $\addrole$.  Function $\addrole(r)$ adds role $r$ to the candidate role hierarchy: inheritance relations involving $r$ are added, and the assigned users and assigned permissions of $r$'s newly acquired ancestors and descendants are adjusted by removing inherited users and permissions\fullonly{, in a similar way as in the construction of the role hierarchy in Phase 3}.  Removing a role $r$ and then restoring $r$ using $\addrole$ leaves the policy unchanged.

\fullonly{When testing whether $\epsilon$-consistency is violated, it is sufficient
to check the size of $T \setminus \mean{\pi}$.  It is unnecessary to
consider $\mean{\pi} \setminus T$, because it is always empty; to see this,
note that $\mean{\pi}$ equals $T$ at the beginning of Phase 4, and Phase 4
only removes roles, which can only decrease $\mean{\pi}$.}


The following auxiliary functions are used in $\removerole$.
isDescendant($r$,$r'$) holds if $r$ is a descendant of $r'$, as determined
by following the parent-child relations in the $\children$ dictionary.  The set of
authorized users of $r$, denoted $\authu(r)$, is the set of users in
$\assignedu(r)$ or $\assignedu(r')$ for some $r'$ senior to $r$;
this is the same as in RBAC.  The notion of authorized permissions must be
defined differently in TRBAC than RBAC, because, with strongly-restricted
inheritance, the inherited permissions of a role $r$ may be associated with
BPESs different than $\ta(r)$.  With strongly-restricted inheritance, the
set of authorized permissions of $r$, denoted $\authp(r)$, is the set of
permission-BPES pairs $\tuple{p,\bpes}$ such that (1) each directly assigned
permission of $r$ is paired with $\ta(r)$ and (2) each permission $p$
inherited by $r$ is paired with the semantic union of the BPESs of the
junior roles from which it is inherited.  With weakly-restricted
inheritance, $\authp(r)$ is the set of permission-BPES pairs
$\tuple{p,\ta(r)}$ such that $p$ is in $\assignedp(r)$ or
$\assignedp(r')$ for some $r'$ junior to $r$; we use a set of pairs for
uniformity with the case of strongly-restricted inheritance.

\begin{figure}[tb]
\fbox{\begin{minipage}[t]{1.0\linewidth}
\begin{tabbing}
$\pi = \mbox{policy from Phase 3}$\\
$q = \qpol(\pi)$\\
$\wklist = \mbox{list of removable roles in $\pi$}$\\
$\changed = \true$\\
\whileloop~$\neg\isempty(\wklist) \land \changed$\\
~~\= sort $\wklist$ in ascending order by $\qrole$\\
\> $\changed = \false$\\
\> \forloop~$r$ in $\wklist$\\
\>~~\= $\removerole(r)$\\
\>\> \com {if $\epsilon$-consistency is violated,}\\
\>\> \com {restore $r$.}\\
\>\> \ifstmt~$|T \setminus \mean{\pi}| > \epsilon$\\
\>\>~~\= $\addrole(r)$\\
\>\>\> $\wklist$.remove($r$)\\
\>\> \elsestmt\\
\>\>\> \com{if policy quality improved,}\\
\>\>\> \com{keep the change.}\\
\>\>\> \ifstmt~$\qpol(\pi) < \delta q$\\
\>\>\>~~\= $\changed = \true$\\
\>\>\>\> $q = \qpol(\pi)$\\
\>\>\>\> $\wklist$.remove($r$)\\
\>\>\> \elsestmt\\
\>\>\>\> \com{undo the change, i.e., restore $r$}\\
\>\>\>\> $\addrole(r)$\\
\\
\end{tabbing}
\end{minipage}}
\fbox{\begin{minipage}[t]{1.0\linewidth}
\begin{tabbing}
\function~\removerole($r$)\\
\forloop~$\parent$ in \parents($r$)\\
~~\= \com{remove $r$ from its parents}\\
\> \children(\parent).remove($r$)\\
\> \forloop~$\child$ in \children($r$)\\
\>~~\=\com{if $\child$ is not a descendant of $\parent$}\\
\>\> \com{after removing $r$, add an inheritance}\\
\>\> \com{edge between $\child$ and $\parent$.}\\
\>\> \ifstmt~$\neg$ isDescendant(\child,\parent)\\
\>\>~~\= \children(\parent).add(\child)\\
\>\>\>	 \parents(\child).add(parent)\\
\> \forloop~$u$ in \assignedu($r$)\\
\>\> \com{if $u$ is not authorized to $\parent$ after}\\
\>\>\com{removing $r$, add $u$ to assigned users}\\
\>\>\com{of \parent.}\\
	\ifstmt~$u\not\in\authu(\parent)$\\
\>\>~~\assignedu($\parent$).add($u$)\\
\forloop~$\child$ in \children($r$)\\
\> \parents(\child).remove($r$)\\
\> \forloop~ $p$ in $\assignedp(r)$\\
\>\> \com{if $p$ is not fully authorized to $\child$}\\
\>\> \com{after removing $r$, add $p$ to assigned}\\
\>\> \com{permissions of $\child$.}\\
\>\> \ifstmt~$\tuple{r,\ta(\child)} \not\in \authp(\child)$\\
\>\> ~~ \assignedp(\child).add($p$)\\
$\rcand$.remove($r$)
\end{tabbing}
\end{minipage}}
\vspace{-1ex}
\caption{Phase 4: Remove roles.}
\label{fig:remove-roles}
\label{fig:removeRole}
\end{figure}



\section{Datasets\vspace{-0.2ex}}
\label{sec:datasets}

We generated two datasets based on real-world ACL policies from HP, described in \cite{ene08fast}, and the high-fit synthetic attribute data for these ACL policies described in \cite{xu12algorithms}; see those references for more information about the ACL policies and attribute data.  Briefly, the ACL policies are named americas\_small, apj, domino, emea, firewall1, firewall2, and healthcare.  The synthetic attribute data is generated pseudorandomly, using statistical distributions based on statistical summaries of some real-world attribute data, to make the synthetic data more realistic.  The number of attributes ranges from 20 to 50, depending on the policy size.  The type of attribute values is unimportant (the only operation used by our algorithm on attribute values is equality), so we simply use natural numbers for the values of all attributes.

As outlined in Section \ref{sec:intro}, for each ACL policy, we mine an RBAC policy from the ACLs and the attribute data using Xu and Stoller's elimination algorithm \cite{xu12algorithms}, and pseudorandomly extend the RBAC policy with temporal information several times to obtain TRBAC policies.  For each ACL policy except americas\_small, we create 30 TRBAC policies.  For americas\_small, which is larger, we create only 10 TRBAC policies, to reduce the running time of the experiments.  We extend the RBAC policies in two ways, using different temporal information.

\myparagraph{Dataset with simple PEs}

A {\it simple PE} is a range of hours (e.g., 9am-5pm) that implicitly repeats every day.\fullonly{ We define the WSC of a simple PE to be 1.}  This dataset uses the same simple PEs as in \cite{mitra15trbac}, namely, $[6, 11]$, $[7, 10]$, $[8, 9]$, $[8, 11]$, $[9, 11]$, $[10, 11]$, $[10, 12]$, $[11, 13]$, $[14, 15]$, $[16, 17]$.  These PEs are designed to cover various relationships between intervals, such as overlapping, consecutive, disjoint, and nested.  We choose the number of PEs in each BPES pseudorandomly using a similar probability distribution as in \cite{mitra15trbac}, namely, $\prob(1)=0.78, \prob(2)=0.2, \prob(3)=0.02$.  We choose the specific PEs in each BPES pseudorandomly using a uniform distribution.

\myparagraph{Dataset with complex PEs}

For this dataset, we use periodic expressions based on a hospital staffing
schedule, based on discussions with
the Director of Timekeeping at Stony Brook University Hospital.  The periodic expressions are not taken directly from the hospital's staffing schedule, but they reflect its general nature.  The schedule does not repeat every week, but rather every few weeks, because weekend duty rotates.
Clinicians may work 3 days/week for 12 hours/day starting at 7am or 7pm, or 5 days/week for 8.5 hours/day starting at 7am, 3pm, or 11pm.\fullonly{ The probabilities of these work schedules are 0.144, 0.094, 0.284, 0.284, and 0.194, respectively.}  We create two instances of each of these five types of work schedules, by pseudorandomly choosing the appropriate number of days of the week in each of the four weeks of a Quadweek\fullonly{, using a uniform distribution}.  Each BPES is based on exactly one of the resulting 10 work schedules.  Multiple PEs are needed to represent work schedules that wrap around calendar units; for example, a 7pm-7am shift is represented using two PEs, with time intervals 7pm-midnight and midnight-7am.  The PEs are based on the following sequence of calendars: $C_1$=Quadweeks, $C_2$=Days, $C_3$=Hours, $C_d$=Hours.  The days in a Quadweek are numbered 1..28.  Including Week in the sequence of calendars is not helpful, because most workers' schedules do not repeat on a weekly basis.\fullonly{ For example, consider a clinician who works 3 days/week for 12 hours/day starting at 7am, working Mon,Wed,Fri during the first and second weeks of a quadweek, and Tue,Thu,Sat during the third and fourth weeks.  Assuming weeks start on Monday, this schedule is represented by the PE [$\all$ $\cdot$ Quadweeks + $\{1,3,5,8,10,12,16,18,20,23,25,27\}$ $\cdot$ Days + $\{8\}$ $\cdot$ Hours $\rhd$ 12 $\cdot$ Hours].}



\section{Evaluation}
\label{sec:eval}


The experimental methodology is outlined in Section \ref{sec:intro}.  All experiments use quality change tolerance $\delta=1.001$ (this value gave the best results for the experiments in \cite{xu12algorithms}), $\epsilon=0$, and $w_i=1$ for all weights in WSC.  The policy quality metric is $\polszinterp$, and the inheritance type is weakly restricted, except where specified otherwise. 


Our Java code and datasets are available at \url{http://www.cs.stonybrook.edu/~stoller/software/}.  Periodic expressions are an abstract data type with two implementations: (1) simple PEs, as defined in Section \ref{sec:datasets}, and implemented as pairs of integers, and (2) (general) PEs, as defined in Section \ref{sec:trbac}, and implemented as arrays of arrays of integers.  
These implementations are used in the experiments in Sections \ref{sec:eval:simple} and \ref{sec:eval:complex}, respectively.
Running times include the cost of an end-to-end correctness check that checks equivalence of the input TUPA and the meaning of the mined TRBAC policy; the average cost is about 7\% of the running time.  The experiments were run on a Lenovo IdeaCentre K430 with a 3.4 GHz Intel Core i7-3770 CPU.

\subsection{Experiments using dataset with simple PEs}
\label{sec:eval:simple}

All experiments on this simple PEs dataset use role intersection cutoff $\ric=1$.

\myparagraph{Comparison of original and mined policies}

Figure \ref{fig:experiments-simple} shows detailed results from experiments on this dataset.  In the column headings,
$\mu$ is mean, $\sigma$ is standard deviation, CI is half-width of 95\% confidence interval using Student's t-distribution, and time is the average running time in minutes:seconds.  There is no standard deviation column for INT, because interpretability is unaffected by the role-time assignment and hence is the same for all TRBAC policies generated by extending the same RBAC policy.  Ignore the last 2 columns for now.  The averages and standard deviations are computed over the TRBAC policies created by extending each RBAC policy.  The WSC of the mined TRBAC policy ranges from about 2\% lower (for healthcare) to about 5\% higher (for firewall1) than the WSC of the original TRBAC policy.  The interpretability of the mined policy ranges from about 40\% lower (for firewall-2) to about 1\% lower (for apj) than the interpretability of the original TRBAC policy.  On average over the seven policies, the WSC is about 0.5\% higher, and the interpretability is about 19\% lower.  Thus, our algorithm succeeds in finding the implicit structure in the TUPA and producing a policy with comparable WSC and better interpretability, on average, than the original TRBAC policy.




\begin{figure*}[tb]
  \centering
\begin{tabular}{|l|r|r|r|r|r|r|r|r|r|l|l|l|}
\hline
 & \multicolumn{3}{c|}{Original Policy} & \multicolumn{6}{c|}{Mined Policy} & \multicolumn{1}{c|}{} & \multicolumn{2}{c|}{Avg $|R|$} \\ 
\cline{2-10}\cline{12-13}
\multicolumn{1}{|c|}{Dataset} & \multicolumn{2}{c|}{WSC} & \multicolumn{1}{c|}{INT} & \multicolumn{3}{c|}{WSC} & \multicolumn{3}{c|}{INT} & \multicolumn{1}{c|}{Time} & \multicolumn{1}{c|}{Our} & \multicolumn{1}{c|}{} \\ 
\cline{2-11}
\multicolumn{1}{|c|}{} & \centercell{$\mu$} & \centercell{$\sigma$} &  & \centercell{$\mu$} & \centercell{$\sigma$} & \centercell{CI} & \centercell{$\mu$} & \centercell{$\sigma$} & \centercell{CI} & \centercell{$\mu$} & \multicolumn{1}{c|}{Alg} & \multicolumn{1}{c|}{GTRM} \\ 
\hline
americas\_small & 6975 & 7.5 & 189 & 7098 & 71 & 27 & 138 & 6 & 2.2 & \multicolumn{1}{r|}{48:42} & \multicolumn{1}{r|}{296} & \multicolumn{1}{r|}{} \\ 
\hline
apj & 4879 & 10.0 & 385 & 4813 & 16 & 5.9 & 384 & 3.4 & 1.3 & \multicolumn{1}{r|}{0:15} & \multicolumn{1}{r|}{470} & \multicolumn{1}{r|}{527} \\ 
\hline
domino & 449 & 2.5 & 23 & 450 & 9 & 3 & 18 & 1.9 & 0.70 & \multicolumn{1}{r|}{0:01} & \multicolumn{1}{r|}{29} & \multicolumn{1}{r|}{40} \\ 
\hline
emea & 3929 & 4.4 & 32 & 4065 & 80 & 30 & 32 & 0 & - & \multicolumn{1}{r|}{0:41} & \multicolumn{1}{r|}{99} & \multicolumn{1}{r|}{115} \\ 
\hline
firewall1 & 1533 & 4.1 & 48 & 1603 & 80 & 30 & 37 & 3.4 & 1.3 & \multicolumn{1}{r|}{1:07} & \multicolumn{1}{r|}{93} & \multicolumn{1}{r|}{130} \\ 
\hline
firewall2 & 960 & 1.4 & 7 & 963 & 7.2 & 2.7 & 4 & 1.2 & 0.44 & \multicolumn{1}{r|}{0:02} & \multicolumn{1}{r|}{12} & \multicolumn{1}{r|}{17} \\ 
\hline
healthcare & 168 & 1.4 & 14 & 165 & 1.6 & 0.6 & 12 & 1.2 & 0.44 & \multicolumn{1}{r|}{0:01} & \multicolumn{1}{r|}{16} & \multicolumn{1}{r|}{25} \\ 
\hline
\end{tabular}
\caption{Results of experiments with simple PEs.}
  \label{fig:experiments-simple}
\end{figure*}

\myparagraph{Comparison of FastMiner and CompleteMiner}

In Phase 1.2 (Intersect roles), instead of the FastMiner approach of computing intersections only for pairs of initial roles, we could instead adopt the CompleteMiner approach of computing intersections for all subsets of initial roles \cite{vaidya06roleminer}.  We ran our algorithm, modified to use CompleterMiner, on our simple PE dataset, omitting emea and americas\_small because of their longer running times.  Figure \ref{fig:experiments-fast-complete-simple} shows the results using FastMiner and CompleteMiner.  Surprisingly, CompleteMiner did not improve policy quality: it increased the average WSC by 4\% on average, ranging from 0.2\% (for firewall2) to 11\% (for domino), and it increased (worsened) the average INT by 10\% on average, ranging from 1\% (for apj) to 19\% (for firewall1).  Although one might expect that generating additional candidate roles would only improve the quality of the final policy, the role selection phase uses imperfect heuristics, so additional candidate roles sometimes lead to decreases in policy quality.  Not surprisingly, CompleteMiner is slower: it increased the average running time by 160\% on average, ranging from 15\% for firewall2 to 201\% for apj.


\begin{figure*}[tb]
\centering
\begin{tabular}{|l|r|r|r|r|r|r|r|r|r|r|r|r|r|r|}
\hline
\multirow{3}{*}{Dataset} & \multicolumn{4}{c|}{WSC} & \multicolumn{4}{c|}{INT} & \multicolumn{2}{c|}{Time} \\ \cline{2-11} 
 & \multicolumn{2}{c|}{CM} & \multicolumn{2}{c|}{FM} & \multicolumn{2}{c|}{CM} & \multicolumn{2}{c|}{FM} & CM & FM \\ \cline{2-11} 
 & \centercell{$\mu$} & \centercell{$\sigma$} & \centercell{$\mu$} & \centercell{$\sigma$} & \centercell{$\mu$} & \centercell{$\sigma$} & \centercell{$\mu$} & \centercell{$\sigma$} & \centercell{$\mu$} & \centercell{$\mu$} \\ \hline
healthcare & 168 & 4 & 165 & 1.6 & 14 & 0.4 & 12 & 1.2 & 0:01 & 0:01 \\ \hline
firewall2 & 966 & 9 & 963 & 7.2 & 5 & 1.0 & 4 & 1.2 & 0:02 & 0:02 \\ \hline
firewall1 & 1661 & 64 & 1603 & 80 & 44 & 3.7 & 37 & 3.4 & 1:13 & 1:07 \\ \hline
domino & 500 & 71 & 450 & 9 & 21 & 1.5 & 18 & 1.9 & 0:01 & 0:01 \\ \hline
apj & 4828 & 21 & 4813 & 16 & 388 & 3.7 & 384 & 3.4 & 0:46 & 0:15 \\ \hline
\end{tabular}%
\caption{Results of experiments with Complete Miner (CM) and Fast Miner (FM).}
  \label{fig:experiments-fast-complete-simple}
\end{figure*}

\myparagraph{Comparison of inheritance types}

We ran our algorithm again on the same dataset with all policies except americas\_small, specifying strongly restricted inheritance for the mined policies.  This caused a significant increase in the WSC of the mined policies.  The percentage increase averages 51\% and ranges from 6\% (for apj) to 105\% (for firewall1 and healthcare).  Intuitively, the reason for the increase is that, with strongly restricted inheritance, the temporal information associated with directly assigned and inherited permissions may be different, and this may prevent removing inherited permissions from a role's directly assigned permissions.  Inheritance type has less effect on the average INT, increasing (worsening) it by about 3\% on average.



\myparagraph{Evaluation of choice of initial roles}

Recall from Section \ref{sec:algorithm} that the definition of permBPES in Figure \ref{fig:generate-init} uses the condition $\bpes \sqsubseteq \bpes'$ in order to include in each initial role the permissions that the user has for a BPES $\bpes'$ that semantically contains $\bpes$.  A more obvious alternative is to require $\bpes = \bpes'$ and thereby include only the permissions that the user has for exactly the same BPES $\bpes$.  Let permBPES$^-$ denote that variant of permBPES.  We evaluated the benefit of using permBPES by running our algorithm, modified to use permBPES$^-$ instead of permBPES, for all policies in the simple PE dataset except the largest one, americas\_small, due to its longer running time.  This change increased the average WSC by 37\% on average, ranging from 13\% (for apj) to 85\% (for healthcare).  It increased (worsened) the average INT by 50\% on average, ranging from 9\% (for apj) to 100\% (for emea).  The average running time decreased by 61\% on average, ranging from 31\% slower (for firewall2) (the only policy for which the modified algorithm was slower) to 94\% faster (for emea).

The policy quality benefit of permBPES over permBPES$^-$ can also be demonstrated with a simple example.  Consider the input TUPA $T = \{ \tuple{u_1, p_1,$ $10am{\rm -}5pm},$ $\langle u_1,$ $p_2, 10am{\rm -}noon\rangle,$ $\tuple{u_1, p_3, noon{\rm -}5pm}\}$.  Our algorithm generates a policy with 2 roles and WSC 8; one role has permissions $\{p_1, p_2\}$ during 10am-noon, and the other role has permissions $\{p_1,p_3\}$ during noon-5pm.  The variant of our algorithm that uses permREB$^-$ instead of permBPES generates a policy with 3 roles, each corresponding to one element of the TUPA, and with WSC 9.  Mitra {\it et al.}'s GTRM algorithm \cite{mitra15trbac} also produces that policy, as expected, since its construction of initial roles is more similar to permBPES$^-$ than permBPES.  Mitra {\it et al.}'s CO-TRAPMP-MVCL algorithm \cite{mitra16trbac} may produce either of these policies, depending on the value of a parameter, namely, the threshold $\theta$ for degree of overlap.

We also evaluated the effect of using both permBPES and permBPES$^-$, \ie, of replacing the call ${\rm permBPES}(u,T)$ with ${\rm permBPES}(u,T) \union {\rm permBPES}^-(u,T)$.  This change increased the average WSC by 0.1\% and the average INT by 0.2\%.  It also increased the average running time by 22\% on average, ranging from 7\% faster (for firewall1) to 60\% slower (for domino).

We considered reducing the cost of Phase 1.1 by removing the first call to addRole.  Note that Mitra {\it et al.}'s algorithm does not include an analogue of this call.  This change increased the average WSC by 9\% on average over the policies used in this experiment (all except americas\_small), ranging from 7\% (for emea and firewall2) to 10\% (for domino).  It increased (worsened) the average INT by 8\% on average over those policies, ranging from 2\% (for firewall2) to 12\% (for firewall1).




\myparagraph{Comparison with Mitra {\it et al.}'s GTRM algorithm}

We ran Mitra {\it et al.}'s GTRM algorithm \cite{mitra15trbac}, and our algorithm with number of roles as policy quality metric (because GTRM algorithm optimizes this metric), on our dataset with simple PEs.  Their code supports only simple PEs, so we used only the simple PE dataset in the comparison.  Their code, implemented in C, gave an error (``malloc: ...: pointer being freed was not allocated'') on some TRBAC policies generated for emea and firewall1; we ignored those results.  Their code did not run correctly on americas\_small, so we omitted it from this comparison.

The last two columns of Figure \ref{fig:experiments-simple} show the numbers of roles generated by the two algorithms.  Standard deviations are omitted to save space but are small: on average, 3\% of the mean, for both algorithms.  The GTRM algorithm produces 34\% more roles than ours, on average.  Our algorithm produces hierarchical policies, and their algorithm produces flat policies, but this does not affect the number of roles.  There are many other differences between the algorithms, discussed in Section \ref{sec:related}, which contribute to the difference in results.  The above paragraph on evaluation of choice of initial roles describes two experiments that explore differences between our algorithm and the GTRM algorithm and quantify the benefit of those differences.  The effects of some other differences between the two algorithms, such as the use of elimination {\it vs.}  selection in Phase 4, were investigated in the untimed case in \cite{xu12algorithms} and likely have a similar impact here.



\subsection{Experiments using dataset with complex PEs}
\label{sec:eval:complex}




\begin{figure*}[tb]
  \centering
\begin{tabular}{|l|r|r|r|r|r|r|r|r|r|l|l|}
\hline
 & \multicolumn{3}{c|}{Original Policy} & \multicolumn{6}{c|}{Mined Policy} & \multicolumn{1}{c|}{} & \multicolumn{1}{c|}{} \\ 
\cline{2-10}
\multicolumn{1}{|c|}{Dataset} & \multicolumn{2}{c|}{WSC} & \multicolumn{1}{c|}{INT} & \multicolumn{3}{c|}{WSC} & \multicolumn{3}{c|}{INT} & \multicolumn{1}{c|}{$\ric$} & \multicolumn{1}{c|}{Time} \\ 
\cline{2-10}\cline{12-12}
\multicolumn{1}{|c|}{} & $\mu$ & $\sigma$ &  & $\mu$ & $\sigma$ & CI & $\mu$ & $\sigma$ & CI & \multicolumn{1}{c|}{} & \multicolumn{1}{c|}{$\mu$} \\ 
\hline
apj & 16836 & 159 & 385 & 16879 & 165 & 205 & 383 & 3.1 & 3.8 & \multicolumn{1}{r|}{1} & \multicolumn{1}{r|}{72:42} \\ 
\hline
domino & 1156 & 49 & 23 & 1256 & 64 & 24 & 16 & 2.0 & 0.7 & \multicolumn{1}{r|}{1} & \multicolumn{1}{r|}{0:34} \\ 
\hline
emea & 5975 & 99 & 32 & 7309 & 354 & 440 & 32 & 0 & 0 & \multicolumn{1}{r|}{0.4} & \multicolumn{1}{r|}{41:24} \\ 
\hline
firewall1 & 3712 & 97 & 48 & 6534 & 509 & 190 & 46.8 & 3.9 & 4.9 & \multicolumn{1}{r|}{0.4} & \multicolumn{1}{r|}{324:48} \\ 
\hline
firewall2 & 1269 & 37 & 7 & 1316 & 56 & 21 & 3.4 & 1.3 & 0.5 & \multicolumn{1}{r|}{1} & \multicolumn{1}{r|}{1:00} \\ 
\hline
healthcare & 560 & 35 & 14 & 592 & 38 & 48 & 8.8 & 1.2 & 1.5 & \multicolumn{1}{r|}{1} & \multicolumn{1}{r|}{11:00} \\ 
\hline
\end{tabular}
  \caption{Results of experiments with complex PEs.}
  \label{fig:experiments-complex}
\end{figure*}

\myparagraph{Comparison of original and mined policies}

Figure \ref{fig:experiments-complex} shows detailed results from experiments on this dataset.  The original TRBAC policies here have higher WSC than the ones in Section \ref{sec:eval:simple}, because complex PEs have higher WSC than simple PEs.  We averaged over 30 TRBAC policies each for domino and firewall2, and (to reduce the running time of the experiments) 5 TRBAC policies each for the others.  For emea and firewall1, we use $\ric=0.4$ instead of $\ric=1$ to reduce the running time.  The average WSC of the mined TRBAC policies ranges from 0.3\% higher (for apj) to 76\% higher (for firewall1) than the WSC of the original TRBAC policy.  The average interpretability of the mined TRBAC policies ranges from 52\% lower (for firewall2) to 0.5\% lower (for apj) than the interpretability of the original TRBAC policy.\fullonly{ On average over the four policies for which we use $\ric=1$, the WSC is 5\% higher, and the interpretability is 30\% lower.  On average over the two policies for which we use $\ric=0.4$, the WSC is 49\% higher, and the interpretability is 1\% lower.}  On average over all six policies, the WSC is 19\% higher, and the interpretability is 20\% lower.  Thus, our algorithm finds most of the implicit structure in the TUPA and produces a policy with moderately higher WSC and better interpretability, on average, than the original TRBAC policy.  The results can be improved by using larger $\ric$, at the expense of higher running time.

The higher running times, compared to the dataset with simple PEs, are due primarily to the larger number of candidate roles created by role intersection (there are more overlaps between BPESs in this dataset), and secondarily to the larger overhead of manipulating more complex PEs.

\myparagraph{Benefit of general PEs}

PEs can be translated into sets of simple PEs. For example, the set of PEs \{[$\all$ $\cdot$ Weeks + \{1,2,7\} $\cdot$ Days + \{1\} $\cdot$ Hours $\rhd$ 8 $\cdot$ Hours]\} can be translated to the set of simple PEs \{[1,9], [25,33], [145,153]\}.  However, PEs are generally more compact and efficient than simple PEs.  For example, in experiments with the healthcare, domino, and firewall2 policies, which have the smallest WSCs among our example policies, using this translation and simple PEs was about 5x, 12x, and 14x slower, respectively, than using general PEs.

\myparagraph{Cost-benefit trade-off from role intersection cutoff}

\begin{figure}[tb]
  \centering
  \includegraphics[width=130mm]{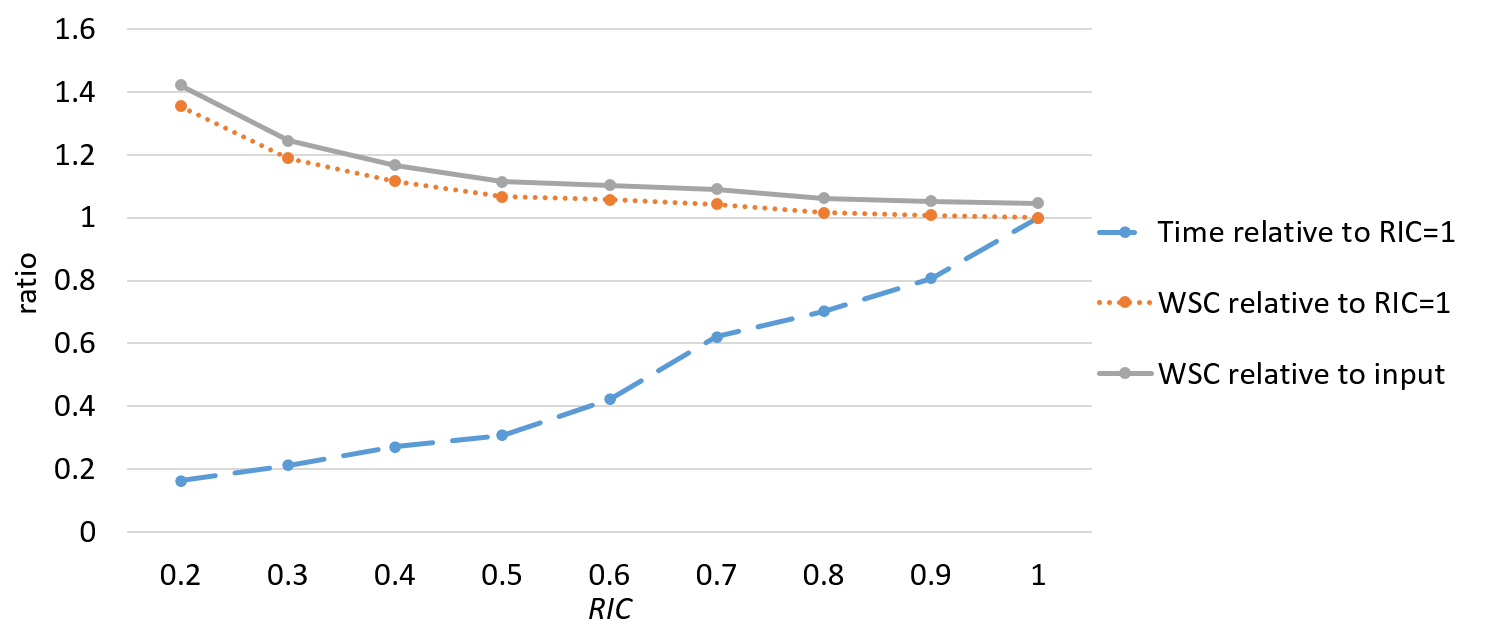}
  \caption{Relative running time and relative WSC as functions of $\ric$.}
  \label{fig:experiments-RIC}
\end{figure}

We investigated the cost-benefit trade-off when varying the role intersection cutoff $\ric$.  Figure \ref{fig:experiments-RIC} shows running time and WSC as functions of $\ric$, averaged over apj, domino, firewall2, healthcare, which are four of the smaller policies.  The trade-off is favorable: as $\ric$ decreases, running time decreases much more rapidly than WSC increases.  For example, at $\ric=0.5$, running time is 70\% lower than with $\ric=1$, and WSC is only 11\% higher.

\myparagraph{Benefit of new $\ric$ metric}

We evaluated the advantage of the userfulness-for-intersection ($\ui$) metric in Section \ref{sec:algorithm} over $\coveredup$, which is the $\ui$ metric in our DBSec 2016 paper \cite{stoller16mining}.  In experiments with apj, domino, emea, firewall2, and healthcare, for $\ric=0.4$, mining with $\coveredup$ as the $\ui$ metric takes 2.5 times longer and produces policies with 17\% higher WSC than mining with the new $\ui$ metric, on average over those policies.  


\subsection{Comparison with Mitra {\it et al.}'s CO-TRAPMP-MVCL algorithm}
\label{sec:eval:CO-TRAP}

Mitra {\it et al.}'s CO-TRAPMP-MVCL algorithm, called the CTR algorithm for brevity, minimizes a variant of WSC, called cumulative overhead of temporal roles and permissions (CO-TRAP), defined by $w_{\ta}.|\ta| + w_{\pa}.|\pa|$, where $w_{\ta}$ and $w_{\pa}$ are user-specified weights \cite{mitra16trbac}.  Mitra {\it et al.} use $w_{\pa}$ = $w_{\ta}$ = 1 for their experiments, and we use the same values.  In these experiments, we run our algorithm with the following weights for WSC: $w_1 = 0$, $w_2 = 0$, $w_3 = 1$, $w_4 = 0$, $w_5 = 1$.  This means the WSC equals $|\pa| + |\ta|$, the same as CO-TRAP.  CO-TRAP is designed for non-hierarchical policies, so we flatten the hierarchical policies produced by our algorithm and then compute CO-TRAP for the flattened policies.  Flattening transforms a hierarchical TRBAC policy into an equivalent non-hierarchical policy, by adding direct user-role assignments for all role memberships that are inherited in the hierarchical policy, and then removing the role hierarchy.  Coincidentally, flattening leaves $\ta$ and $\pa$ unchanged, so we get the same result regardless of whether we compute CO-TRAP for the hierarchical policy or the flattened policy.


\newcommand{\cti}{{\it CTI}}
\newcommand{\oti}{{\it OTI}}

\myparagraph{Dataset}

Our experimental comparison with the CTR algorithm uses the datasets generated by Mitra {\it et al.} for their experiments with CTR algorithm described in \cite{mitra16trbac}.  It is based on the same real-world ACL policies from HP as our datasets described in Section \ref{sec:datasets}.  It contains TRBAC policies generated by mining non-temporal RBAC policies using Ene {\it et al.}'s algorithm \cite{ene08fast}, and then extending them with synthetic temporal information containing simple PEs.  First, they create 10 sets of contained time intervals (the intervals in each set are totally ordered by the subset relation) and 10 sets of overlapping time intervals (every pair of intervals in each set has a non-empty intersection).  They create a role-time assignment by pseudorandomly associating some of these time intervals with each role, selecting from the sets of contained time intervals and overlapping time intervals with probability $d$ and $1-d$, respectively, where $d$ is a parameter of the generation process.  They generate five datasets, each for a different value of $d$: 1, 0.75, 0.50, 0.25, and 0.  These datasets are denoted c100, c75o25, c50o50, c25o75, and o100, respectively.  Each dataset contains 30 TRBAC policies with different pseudorandom role-time assignments.

\myparagraph{Results}

Figure \ref{fig:compare-ctr} shows the average $\mu$ and standard deviation $\sigma$ of CO-TRAP for policies generated by our algorithm, and average CO-TRAP for policies generated by CTR algorithm as reported in \cite[Table 6]{mitra15trbac}.  The average CO-TRAP for policies generated by our algorithm ranges from 68\% lower (for healthcare c100) to 19\% lower (for emea o100) than the corresponding results for the CTR algorithm.  On average over all five datasets for all eight ACL policies, results for policies generated by our algorithm are 41\% lower than results for policies generated by the CTR algorithm.  Thus, our algorithm is significantly more effective than the CTR algorithm at minimizing CO-TRAP.


It took less than 2 minutes to run our algorithm for all 30 TRBAC policies generated from each of the ACL policies healthcare, domino, firewall2, and emea.  It took less than 2 minutes to run our algorithm for each TRBAC policy generated from apj, firewall1, and americas\_large (an ACL policy from HP not used in the datasets described in Section \ref{sec:datasets}).  It took approximately 24 minutes to run experiments for each TRBAC policy generated from americas\_small.  Mitra {\it et al.} report that ``each individual run took no more than 24 minutes'' \cite{mitra16trbac}.  Although these measurements are from experiments on different hardware and software platforms (our algorithm is implemented in Java, and CTR algorithm is implemented in C), they suggest that running times of our algorithm and CTR algorithm are comparable.


\begin{figure}
\centering
\resizebox{\textwidth}{!}{%
\begin{tabular}{|l|l|l|l|l|l|l|l|l|l|l|l|l|l|l|l|}
\hline
\multirow{3}{*}{Dataset} & \multicolumn{3}{c|}{c25\_o75} & \multicolumn{3}{c|}{c50\_o50} & \multicolumn{3}{c|}{c75\_o25} & \multicolumn{3}{c|}{c100} & \multicolumn{3}{c|}{o100} \\ \cline{2-16} 
 & \multicolumn{2}{c|}{Our Algorithm} & \multicolumn{1}{c|}{CTR} & \multicolumn{2}{c|}{Our Algorithm} & \multicolumn{1}{c|}{CTR} & \multicolumn{2}{c|}{Our Algorithm} & \multicolumn{1}{c|}{CTR} & \multicolumn{2}{c|}{Our Algorithm} & \multicolumn{1}{c|}{CTR} & \multicolumn{2}{c|}{Our Algorithm} & \multicolumn{1}{c|}{CTR} \\ \cline{2-16} 
 & \multicolumn{1}{c|}{$\mu$} & \multicolumn{1}{c|}{$\sigma$} & \multicolumn{1}{c|}{$\mu$} & \multicolumn{1}{c|}{$\mu$} & \multicolumn{1}{c|}{$\sigma$} & \multicolumn{1}{c|}{$\mu$} & \multicolumn{1}{c|}{$\mu$} & \multicolumn{1}{c|}{$\sigma$} & \multicolumn{1}{c|}{$\mu$} & \multicolumn{1}{c|}{$\mu$} & \multicolumn{1}{c|}{$\sigma$} & \multicolumn{1}{c|}{$\mu$} & \multicolumn{1}{c|}{$\mu$} & \multicolumn{1}{c|}{$\sigma$} & \multicolumn{1}{c|}{$\mu$} \\ \hline
healthcare & 92 & 17 & 279 & 124 & 33 & 287 & 142 & 39 & 281 & 83 & 13 & 265 & 191 & 40 & 283 \\ \hline
domino & 420 & 43 & 627 & 391 & 25 & 631 & 402 & 37 & 632 & 405 & 35 & 625 & 405 & 38 & 634 \\ \hline
apj & 1813 & 11 & 2375 & 1817 & 9 & 2524 & 1832 & 9 & 2605 & 1799 & 8 & 2303 & 1849 & 5 & 2640 \\ \hline
firewall1 & 1109 & 120 & 2819 & 1142 & 88 & 3370 & 1202 & 110 & 3432 & 1076 & 89 & 2704 & 1276 & 94 & 3353 \\ \hline
firewall2 & 602 & 1 & 941 & 612 & 48 & 941 & 603 & 2 & 941 & 602 & 2 & 947 & 604 & 5 & 944 \\ \hline
emea & 5542 & 192 & 7245 & 5634 & 201 & 7245 & 5751 & 193 & 7245 & 5385 & 176 & 7245 & 5856 & 159 & 7245 \\ \hline
americas\_large & 67288 & 964 & 94515 & 69077 & 795 & 96020 & 68108 & 878 & 96797 & 60734 & 1029 & 91971 & 62393 & 1012 & 97110 \\ \hline
americas\_small & 3616 & 247 & 9563 & 3834 & 213 & 10052 & 4358 & 264 & 10446 & 3321 & 180 & 8567 & 4296 & 228 & 10618 \\ \hline
\end{tabular}%
}
\caption{Comparison of our algorithm and the CO-TRAPMP-MVCL (a.k.a. CTR) algorithm using the CO-TRAP metric.}
\label{fig:compare-ctr}
\end{figure}

\section{Related Work}
\label{sec:related}

We discuss related work on TRBAC policy mining and then related work on RBAC mining.  Role mining (for RBAC or TRBAC) is also reminiscent of some other data mining problems, but algorithms for those other problems are not well suited to role mining.  For example, association rule mining algorithms are designed to find rules that are probabilistic in nature\fullonly{ and are supported by statistically strong evidence}. They are not designed to produce a set of rules strictly consistent with the input that completely covers the input and is minimum-sized among such sets of rules.

\subsection{Related Work on TRBAC Policy Mining}

Mitra {\it et al.} define a version of the TRBAC policy mining problem, called the generalized temporal role mining (GTRM) problem, based on minimizing the number of roles.  They present an algorithm, which we call the GTRM algorithm, for approximately solving this problem \cite{mitra15trbac}.  It is an improved version of their earlier work \cite{mitra13toward}.

Mitra {\it et al.} also define another version of the TRBAC policy mining problem, called cumulative overhead of temporal roles and permissions minimization problem (CO-TRAPMP), based on minimizing the CO-TRAP metric described in Section \ref{sec:eval:simple}.  They present another algorithm, called CO-TRAPMP-MVCL, for heuristically solving this problem \cite{mitra16trbac}.

Our algorithm is more flexible than the GTRM and CO-TRAPMP-MVCL algorithms, because our algorithm can optimize a variety of metrics, including WSC and interpretability.  The importance of interpretability is discussed in Section \ref{sec:intro}.  WSC is a more general measure of policy size than number of roles or CO-TRAP and can more accurately reflect expected administrative cost.  For example, the average number of role assignments per user is a measure of expected administrative effort for adding a new user \cite{uzun15migrating}, and this can be reflected in WSC by giving appropriate weight to the size of the user-role assignment.  Neither number of roles nor CO-TRAP take the size of the user-role assignment into account.



Our algorithm produces hierarchical TRBAC policies.  The GTRM and CO-TRAPMP-MVCL algorithms produce flat TRBAC policies.  Role hierarchy is a well-known feature of RBAC that can significantly reduce policy size and administrative effort by avoiding redundancy in the policy.

Our algorithm and the GTRM algorithm have a similar high-level structure: they both (1) create a large set of candidate roles based on the input TUPA, (2) merge some candidate roles, and then (3) select a subset of the candidate roles to include in the final policy.  The algorithms also have many differences.  Some differences are related to policy quality metric and role hierarchy, as discussed above.  Some other differences are: (1) Our algorithm determines which candidate roles to include in the final policy by elimination of low-quality roles, instead of selection of high-quality roles.  We showed that elimination gives better results in the untimed case \cite{xu12algorithms}.
(2) Our algorithm creates more initial roles than the GTRM algorithm.  The benefit of creating these additional initial roles is shown in Section \ref{sec:eval:simple}.  The GTRM algorithm creates {\em unit roles}, which are similar to our initial roles but have only one permission.  In particular, an initial role created by the second call to addRole in our algorithm is a unit role only when $P$ is a singleton set and ${\rm permBPES}(u,T) = {\rm permBPES}^-(u,T)$; we not expect this to be a common case, since most temporal roles have multiple permissions.
(3)\shortonly{ Our algorithm performs fewer types of role intersections than theirs.  Specifically, it omits types of role intersections that  create PEs with time intervals that do not appear in the input, since these PEs are probably not natural (intuitive) ones in the application domain.}
\fullonly{ Our algorithm performs fewer types of intersections than the GTRM algorithm.  The GTRM algorithm performs five types of intersections, corresponding to $r_a, r_b, r_c, r_d, r_e$ in \cite[Algorithm 1]{mitra15trbac}.  Our algorithm performs only intersections corresponding to $r_a$.  We omit $r_b$ and $r_c$ because they may create PEs with time intervals that do not appear in the input TUPA and are not intuitive to security administrators.
  We omit $r_d$ and $r_e$ because Phase 3 would merge those roles back into the roles from which they were created.}
\fullonly{(4) Our algorithm performs more merges; specifically, the GTRM algorithm does not include case (2a) of the merge in Phase 2 of our algorithm.}

The CO-TRAPMP-MVCL algorithm has a different high-level structure than our algorithm: roughly speaking, it (1) repeatedly generates a small set of candidate roles based on the current set of uncovered triples and adds the best one among them to the policy, and then (2) merges some roles.  In the experiments in Section \ref{sec:eval:CO-TRAP}, our algorithm produces higher-quality policies than CO-TRAPMP-MVCL algorithm, as measured using the CO-TRAP metric which the CO-TRAPMP-MVCL algorithm is designed to optimize.

Our implementation supports periodic expressions for specifying temporal information, while Mitra {\it et al.}'s implementations of the GTRM and CO-TRAPMP-MVCL algorithms support only ranges of hours that implicitly repeat every day.  Design and implementation of operations on sets of PEs is non-trivial.\fullonly{ This includes operations such as testing whether one set of PEs covers all of the time instants covered by another set of PEs, and handling numerous corner cases, such as time intervals that wrap around calendar units (e.g., a 7pm-7am work shift).}

\subsection{Related Work on RBAC Mining}

A survey of work on RBAC mining appears in \cite{hachana12role}.  The most closely related work is Xu and Stoller's elimination algorithm \cite{xu12algorithms}.  We chose it as the starting point for design of our algorithm, because in the experiments in \cite{xu12algorithms}, it optimizes WSC more effectively than Hierarchical Miner \cite{molloy10mining}\fullonly{ and the Graph Optimisation role mining algorithm \cite{zhang07role}}, while simultaneously achieving good interpretability, and it optimizes WSCA, an interpretability metric defined in \cite{molloy10mining}, more effectively than Attribute Miner \cite{molloy10mining}.

Our algorithm retains the overall structure of the elimination algorithm but differs in several ways, due to the complexities created by considering time.  Our algorithm introduces more kinds of candidate roles than the elimination algorithm, because it needs to consider grouping permissions that are enabled for the same time or a subset of the time of other permissions.  Our algorithm attempts to merge candidate roles; the elimination algorithm does not.  Construction of the role hierarchy is significantly more complicated than in the elimination algorithm; for example, with strongly restricted inheritance, a permission $p$ can be inherited by a role $r$ from multiple junior roles with different BPESs, which may together cover all or only part of the time that $p$ is available in $r$.  This also complicates adjustment of the role hierarchy when removing candidate roles.  The role quality metric used to select roles for removal is more complicated, to give preference to roles that cover permissions for more times.



\lncsonly{\myparagraph{Acknowledgements}}
\articleonly{\myparagraph{Acknowledgements}}
\acmonly{\section{Acknowledgements}}

We thank the authors of \cite{mitra15trbac,mitra16trbac}---Barsha Mitra, Shamik Sural, Vijayalakshmi Atluri, and Jaideep Vaidya---for sharing their code and datasets with us and helping us understand their work.

\bibliographystyle{abbrv}
\bibliography{references}

\end{document}
